
\magnification=\magstep1
\hsize = 32 pc
\vsize = 42 pc
\baselineskip = 24 true pt
\def\vs{\vskip 0.5 true cm}

\hfill IP-BBSR/93-65 \break

\hfill October-93 \break
\vs
\vs
\centerline {\bf Supersymmetry, Shape Invariance and Exactly}
\centerline {\bf Solvable Noncentral Potentials}
\vs
\def\nd{\noindent}
\nd\centerline {\bf Avinash Khare}

\nd\centerline {Institute of Physics, Sachivalaya Marg}

\nd\centerline {Bhubaneswar -751005, India.}
\vs
\nd\centerline {\bf Rajat K. Bhaduri}

\nd\centerline {Department of Physics and Astronomy}

\nd\centerline {McMaster University, Hamilton, Ontario, Canada L8S 4M1.}
\vs
\nd {\bf Abstract}

Using the ideas of supersymmetry and shape invariance we show that
the eigenvalues and eigenfunctions of a wide class of noncentral
potentials can be obtained in a closed form by the operator method.
This generalization considerably extends the list of exactly solvable
potentials for which the solution can be obtained algebraically in  a
simple and elegant manner. As an illustration, we discuss in detail
the example of the potential
$$V(r,\theta,\phi)={\omega^2\over 4}r^2 + {\delta\over r^2}+{C\over r^2
sin^2\theta}+{D\over r^2 cos^2\theta} + {F\over r^2 sin^2\theta\, sin^2
\alpha\phi} +{G\over r^2 sin^2\theta\, cos^2\alpha\phi}$$
with 7 parameters.Other algebraically solvable examples are also given.

\vfill

\eject

\nd {\bf I. Introduction}

Most text books on nonrelativistic quantum mechanics show how the
harmonic oscillator problem may be elegantly solved by the raising
and lowering operator method [1]. Sometime ago, using the ideas of
supersymmetry [2] and shape invariant potentials (SIP) [3], a
generalization of this method [4] was used to handle many more potentials of
interest. It was found there that this enlarged list included  essentially all
the solvable potentials found in  most text books on nonrelativistic quantum
mechanics. It turns out
that most of these potentials are either one dimensional or are
central potentials which again are essentially one dimensional but on
the half-line. It is natural to enquire if one can also solve some
noncentral potentials in three dimensions by the operator method .
The purpose of this paper is to show that a large number of noncentral
potentials that are  separable in their coordinates
and may have upto  seven parameters can be
solved by the operator method by using the known results for the various
SIP. More precisely, we show that the energy eigenvalues  and eigenfunctions of
noncentral but separable potentials in spherical coordinates $r,\theta,\phi$
can be simply written down by applying supersymmetry and shape invariance
successively to the $\phi,\theta$ and $r$ dependent potentials.

The  procedure outlined here is elegant, and simple,
so that a student of quantum mechanics should be able to appreciate it. Indeed,
we feel that the material presented here, along with that given in ref [4],
may be profitably included in a graduate quantum mechanics course.
Accordingly, we have kept this article at a pedagogical level and
made it  self-contained . In sec.II , we give a brief
review of the main ideas of supersymmetric quantum mechanics and the
concept of shape invariance, and then show how to obtain
the energy eigenvalues and the eigenfunctions of such potentials by a
generalized operator method. Sec.III is the core of this
article. In it we explain  how the results for the known SIP may
be  used to algebraically obtain in a closed form the eigenvalues and the
eigenfunctions for noncentral but separable potentials . As an illustration we
discuss in detail the noncentral potential
$$V_1(r,\theta,\phi)={\omega^2\over 4}r^2 +{\delta\over r^2}+{C\over r^2 {\rm
sin}^2
\theta} +{D\over r^2{\rm cos}^2\theta} +{F\over r^2{\rm sin}^2 \theta\,{\rm
sin}^2 \alpha\phi}+{G\over r^2{\rm sin}^2 \theta\, {\rm cos}^2
\alpha\phi},\eqno{(1.1)}$$
where$\omega$,$\delta$, C,D, G,F and $\alpha$ are arbitrary parameters, and
show that the corresponding energy eigenvalues are
$${E^{(1)}}_{n,n_1,n_2}=\,\biggl[(2n_2+1)+(\delta+l_1^2)^{1\over
2}\biggr]\omega ,\eqno{(1.2)}$$
where
$$l_1^2=\biggl[(2n_1+1)+\sqrt{D+{1\over4}}+\bigg
\{C+\biggl(\sqrt{F+{\alpha^2\over 4}}+\sqrt{G+{\alpha^2\over 4}}+\alpha
(2n+1)\biggr)^2\bigg \}^{1\over 2}\biggr]^2\,.\eqno{(1.3)}$$
The expressions for the energy eigenvalues and the eigenfunctions
of the other noncentral potentials are also given.
\vs
\nd {\bf II. Supersymmetry and Shape Invariance}

Consider a potential $V_-(x)$ whose ground state wave function
$\psi^{(-)}_0=\psi_0(x)$ is known and whose ground state energy has
been adjusted (by adding a suitable constant to $V_-(x)$) to
$E^{(-)}_0=0$. Then the Schr\"odinger equation for the ground state is
(we shall use $\hbar=2M=1$ throughout this article)

$$H_-\psi_0=(-{d^2\over dx^2}+V_-(x))\psi_0 =\,0\;,\eqno{(2.1)}$$
so that
$$V_-(x) = {\psi_0^{''}(x)\over \psi_0(x)}\,.\eqno{(2.2)}$$
If we now define the operators
$$A = {d\over dx} + W(x)\,,\; A^+ = -{d\over dx} + W (x)\,,\eqno{(2.3)}$$
where the superpotential W(x) is related to $\psi_0$ by
$$W(x)= -{\psi'_0(x)\over \psi_0(x)}\,;\; \psi_0(x) = N\, exp [-\int^x W(y)
dy]\,,\eqno{(2.4)}$$
then one finds that
$$H_-\equiv A^+ A =  -{d^2\over dx^2}+W^2 - W'(x)\,.\eqno{(2.5)}$$
The supersymmetric partner potential $H_+$ is given by
$$H_+\equiv AA^+ = - {d^2\over dx^2}+W^2+W'(x)\,.\eqno{(2.6)}$$
If $\psi^{(-)}_n$ and $\psi^{(+)}_n$ denote the eigenfunctions of the
Hamiltonian $H_-$ and $H_+$ with eigenvalue $E^{(-)}_n$ and
$E^{(+)}_n$ respectively then it is easily shown that
$$ E^{(+)}_n= E^{(-)}_{n+1}\,,\eqno{(2.7)}$$
$$\psi^{(+)}_n = (E^{(-)}_{n+1})^{-1/2} A \psi^{(-)}_{n+1}\,,\eqno{(2.8)}$$
$$\psi^{(-)}_{n+1} = (E^{(+)}_n)^{-1/2} A^+ \psi^{(+)}_n\,.\eqno{(2.9)}$$
Further, the reflection and transmission coefficients (or phase
shifts) of the two potentials are also related [2].

The underlying reason for this relationship  is the fact that there
is a supersymmetry in the problem. In particular it is easily shown
that the supercharges Q and $Q^+$ defined by
$$ Q = \pmatrix{0 & 0\cr A & 0\cr}, Q^+ = \pmatrix{0 & A^+\cr 0  &
0\cr}\,, \eqno{(2.10)}$$
along with the Hamiltonian H given by
$$H = \pmatrix{H_- & 0\cr 0 & H_+\cr}\,,\eqno{(2.11)}$$
satisfy the algebra
$$Q^2 = 0 = Q^{+2}\;,[ H, Q ] = 0= [H,Q^+]\, \eqno{(2.12)}$$
$$QQ^+ + Q^+Q = H\,.\eqno{(2.13)}$$

It must be emphasized here that the supersymmetry only gives the
relationship between the eigenvalues and eigenfunctions of the two
Hamiltonians but does not yield  the actual spectrum. For that
purpose one needs an extra integrability condition called the shape
invariance condition which was introduced by Gendenshtein [3]. Stated simply,
if the pair of supersymmetric partner potentials $V_{\pm}(x)$ defined
above are similar in shape and differ only in the parameters which
appear in them, then they are said to be shape invariant. More
specifically, if $V_{\pm}(x;a_0)$ satisfy the requirement
$$V_+(x;a_0) = V_-(x;a_1) + R(a_1)\,,\eqno{(2.14)}$$
where $a_0$ is a set of parameters and $a_1$ is an arbitrary function
of $a_0(a_1 = f(a_0)$) and the reminder $R(a_1)$ is independent of x,
then $V_{\pm}(x;a_0)$ are said to be shape invariant. In such a case
the energy eigenvalue spectrum of the Hamiltonian $ H_{-}$ is given by[3]

$$E^{(-)}_n\,(a_0) = \sum^n_{k=1} R(a_k);\; E^{(-)}_0\,(a_0) =
0\,,\eqno{(2.15)}$$
with $ a_k=f^{k}(a_0),$ i.e., the function f applied k times.
Subsequently, Dutt et al [5] showed that the eigenfunctions
$\psi^{(-)}_n$ of $H_-$ can also be written down algebraically
$$\psi^{(-)}_n(x;a_0) = A^+ (x;a_0) A^+(x;a_1)...
A^+(x;a_{n-1})\psi^{(-)}_0(x;a_n)\eqno{(2.16)}$$
which is clearly a generalization of the operator method for the
celebrated harmonic oscillator potential. Later on Dabrowska et al
[6] algebraically obtained the explicit expressions for
$\psi^{(-)}_n$ for all the known SIP by using Eq.(2.16).

It turned out [7] that there are twelve SIP for which $a_1 =
a_0+\alpha;\alpha$ being a constant. Out of these , eleven  are contained in
the table in ref.[4], while the remaining one was pointed out by Levai
[8].Recently it has been realized [9] that out of these twelve, two are
not really independent, so that one really has ten independent SIP (with
$a_1=a_0+\alpha\;$), and for all of which $E_n $ and $\psi_n$ can be
obtained algebraically. Most of these potentials are also contained in the
factorization approach of Schr\"odinger [10]. It is worth repeating that all
the SIP are either
one dimensional or central potentials which are again essentially one
dimensional in nature.
\vs
\nd {\bf III. Shape Invariance and Noncentral Potentials}

Noncentral potentials are normally not discussed in most text books
on  quantum mechanics. This is presumably because most of them are not
analytically solvable. However, it is worth noting that there is a
class of noncentral  potentials in 3-dimensions [11] for which the
Schr\"odinger equation is separable.In spherical coordinates $r,\theta,\phi\; (
0 \leq r \leq \infty,0 \leq \theta \leq \pi, 0 \leq \phi < 2\pi)$ , the most
general
potential for which the Schrodinger equation is separable is given by
$$V(r,\theta,\phi) = V(r) +{V(\theta)\over r^2} + {V(\phi)\over r^2
{\rm sin}^2\theta}\;,\eqno{(3.1)}$$
where $V(r), V(\theta)$ and $V(\phi)$ are arbitrary functions of their
argument. Although we use the same symbol $V$ for simplicity of notation,the
functions $V(r) , V(\theta)$ , and $ V(\phi) $ need not be the same. The
crucial point is that for each of $V(r), V(\theta)$ and $V(\phi)$ one can
choose a well known SIP and
hence can immediately write down the corresponding energy eigenvalues
and eigenfunctions algebraically. Thus by applying supersymmetry  and shape
invariance successively, i.e. first to $V(\phi)$, then to $ V(\theta)$ and
finally to $V(r)$, we can algebraically solve a large class of noncentral
potentials.First , let us see why the Schr\"odinger equation with a potntial of
the form given by Eq.(3.1) is   separable in the $(r,\theta,\phi)$
coordinates.The equation for the wave function $\Psi(r ,\theta, \phi )$ is
$$\bigg [ -({\partial^2\Psi\over\partial^2r} + {2\over r} {\partial\Psi\over
\partial r })-{1\over
r^2} ({\partial^2\Psi\over \partial\theta^2}+{\rm cot} \theta
{\partial\Psi\over \partial\theta}) - {1\over r^2 {\rm  sin}^2\theta}
{d^2\Psi\over \partial\phi^2}\bigg ] = (E - V) \Psi\,.\eqno{(3.2)}$$
It is convenient to write  $\Psi(r,\theta,\phi)$  as
$$\Psi(r,\theta,\phi) = {R(r)\over r} {H(\theta)\over ({\rm sin}
\theta)^{1/2}} K(\phi)\,.\eqno{(3.3)}$$
Substituting Eq.(3.3) in Eq.(3.2), we obtain
$$- {1\over R}{d^2 R\over d r^2}+  V(r)-{1\over {4 r^2}}
+{1\over r^2}[-{1\over H}{d^2 H\over d\theta^2}+
V(\theta)-{1\over 4}{\rm  cosec}^2 \theta] +{1\over r^2{\rm
sin}^2\theta}[-{1\over K}{d^2 K\over d\phi^2} + V( \phi)] = E\;.\eqno{(3.4)}$$
Suppose $K( \phi )$ obeys the equation
$$- {d^2 K\over d \phi^2} + V(\phi) K(\phi) = m^2 K(\phi)\; .\eqno(3.5)$$
Substituting this in Eq.(3.4), the $\phi$-dependence in it is eliminated,
resulting in the equation,
$$-{1\over R} {d^2 R\over d r^2}+ [ V(r) - {1\over 4 r^2} ]
+ {1\over r^2} [ - {1\over H}{d^2 H\over d \theta^2} + V(\theta) + ( m^2 -{
1\over 4}) {\rm cosec}^2 \theta ] = E \;.\eqno(3.6)$$
Next, let $H (\theta ) $ obey the equation
$$- {d^2 H\over d \theta^2} + [ V(\theta) + ( m^2 - {1\over 4} )\,{\rm cosec}^2
 \theta]\, H(\theta) = l^2 H( \theta )\;.\eqno(3.7)$$
A further substitution of this equation in Eq.(3.6) then completes the
separation of the variables , giving the radial equation ,
$$- {d^2 R\over d r^2} + [ V(r) +{ ( l^2 - {1\over 4} )\over r^2}]\, R(r)
 = E\, R(r) .\eqno(3.8)$$
The three solutions given by Eqs. (3.5), (3.7), and (3.8) may be implemented
algebraically by using the well-known results for the SIP in each case.

As one example, we discuss the potential given by Eq.(1.1) in some detail.
Comparing Eqs. (1.1) and (3.1) , we obtain
$$V_1(\phi) = F\, {\rm cosec}^2 \alpha \phi + G\, {\rm sec}^2 \alpha
\phi\;,\eqno(3.9)$$
where the subscript 1 refers to the particular example under cosideration.
Now, we know from [4] that if the superpotential $W$ is chosen as
$$ W = A\,{\rm tan}\, \alpha\phi - B\,{\rm  cot}\, \alpha
\phi\;,\eqno{(3.10)}$$
then, using Eqs.(2.5) and (2.6) , the corresponding (shape invariant)
 potentials $V_{\mp}$ are given by
$$V_{\mp}(\phi) = - (A+B)^2+A(A\mp\alpha)\,{\rm sec}^2\alpha\phi +
B(B\mp\alpha)\,{\rm  cosec}^2\alpha\phi\;.\eqno{(3.11)}$$
Hence,
$$V_{+}(A,B,\alpha,x)=V_{-}(A+\alpha,B+\alpha,\alpha,x)+(A+B+2\alpha)^2
-(A+B)^2\;.\eqno(3.12)$$
On comparing Eqs.(3.9) and (3.11) we see that , discounting the overall
 constant \break \hfil $-(A+B)^2$, $V_1$ and $V_{-}$are identical if
$$ F = B\,(B - \alpha) ,\,\; G = A\,(A - \alpha)\;.$$
Thus the energy eigenvalues $m^2_1$ of Eq.(3.5) with $V(\phi)=V_1(\phi)$
follow from Eqs.(2.14),\break\hfil (2.15) and (3.12) ,
$$m^2_1 = (A + B + 2n \alpha)^2\;,\;\;\;\;n\,=\,0,1,2,... \eqno{(3.13)}$$
The corresponding eigenfunctions are given by [4] ,
$$K_1(\phi) = ({\rm sin}\, \alpha\phi)^{B/\alpha}\, ({\rm
cos}\,\alpha\phi)^{A/\alpha}\,P^{{B\over \alpha}-{1\over 2}, {A\over
\alpha}-{1\over 2}}_{n}\,({\rm cos}\, 2\alpha\phi)\;.\eqno{(3.14)}$$

Having solved the eigenvalue equation for the $\phi-$part, we
proceed to solve the Schr\"odinger equation given by Eq.(3.7) for the
$\theta$-vavariable .Again , comparing Eqs.(1.1) and (3.1), we see that
$$V_1(\theta)\,=\,C\,{\rm cosec}^2\,\theta\,+\,D\,{\rm
sec}^2\,\theta\;.\eqno(3.15)$$
The resulting Schr\"odinger equation for $H_1(\theta)$ is
$$- {d^2H_1\over d\theta^2} +[\,(C+m^2_1-1/4)\,
{\rm cosec}^2\theta + D\,{\rm  sec}^2 \theta\,]\,H_1 = l^2_1\,H_1\,.\eqno{(3.16
)}$$
Using the same algebraic procedure as before ,  the
eigenvalues $l^2_1$ and the eigenfunctions $H_1(\theta)$ of  Eq.(3.16)
are found to be
$$l^2_1 = (A_1+B_1+2n_1+{1\over 2})^2\,,\eqno{(3.17)}$$
$$H_1(\theta) = ({\rm sin}\,\theta)^{B_1+1/2} ({\rm cos}\,\theta)^{A_1}
\,P^{B_1,A_1-{1\over 2}}_{n_1}({\rm cos}\,2\,\theta)\,,\eqno{(3.18)}$$
where
$$ D = A_1 (A_1 -1)\,,\; C + m^2_1 = B^2_1\;.\eqno{(3.19)}$$
Notice that in general there is no degeneracy in $l_1$ except in the
special case when C = 0 , and $\alpha$ is a rational number. In particular for
C = 0 , $l^2_1$ is given by
$$l^2_1 =
1+2n_1+(2n+1)\alpha+\sqrt{D+1/4}+\sqrt{F+\alpha^2/4}+\sqrt{G+\alpha^2/
4}\;.\eqno{(3.20)}$$
So if $\alpha = p/q$ , where $ p,q$  are integers with no common factors , then
there is degeneracy.

Finally, the radial Eq.(3.8) for the potential (1.1) is given by
$$ -{d^2R_1\over d r^2}+({\omega^2r^2\over
4}+{\delta+l^2_1-1/4\over r^2})R_1=E^{(1)}R_1\;,\eqno{(3.21)}$$
where $E^{(1)}$ denotes the energy eigenvalue . Choosing the superpotential
$$ W = A_2 r - {(B_2+1)\over r}\,,\eqno{(3.22)}$$
the corresponding shape invariant potential $V_-$ is given by
$$V_-(r) = A^2_2 r^2 + {B_2\over r^2}(B_2+1) - A_2(2B_2+3)\;.\eqno{(3.23)}$$
A little algebra will show that
$$V_{+}(A_2,B_2,r)=V_{-}(A_2,B_2+1,r)-A_2(2B_2+3)+A_2(2B_2+1)\;.$$

On comparing Eqs.(3.21) and (3.23) , we see that , apart from the overall
constant $- A_2(2B_2+3)$ , the radial potential in Eq.(3.21) is the same as
$V_{-}$ if
$${\omega^2\over 4}\,=\, A^2_2,\;\;\; \delta+ l_1^2 =
(B_2+1/2)^2\;.\eqno{(3.24)}$$
The energy eigenvalues of the noncentral potential (1.1) may again be
immediately written down as before . We find
$$ E^{(1)}_{n,n_1,n_2} = (2n_2+B_2+{3\over 2})\,\omega\;,\eqno{(3.25)}$$
while the corresponding radial part of the eigenfunction is given by
$$R_1(r) = r^{B_2+1} e^{-\omega r^2/4}\, L^{B_2+{1\over 2}}_{n_2}({1\over 2}
\omega r^2)\;.\eqno{(3.26)}$$
Thus the total eigenfunction for the noncentral potential (1.1) is given by
the form (3.3), with $R_1(r)$, $H_1(\theta)$ and $K_1(\phi)$ given by
Eqs.(3.26) , (3.18) and (3.14) respectively.

In terms of the seven parameters, the energy eigenvalues are given
by Eq.(1.2) $(n,n_1,n_2 = 0,1,2,...)\;,$
where we see the highly nontrivial dependence of $E^{(1)}$ on these
parameters. Note that there is a degeneracy in the problem when
either $\delta$ and/or C are zero and $\alpha$ is a rational number. The
maximum degeneracy is obtained in
case both $\delta$ and C are zero and $\alpha$ is a rational number and
then the energy eigenvalues are given by
$$E^{(1)}_{n,n_1,n_2}  = \bigg \{
2(n_1+n_2+n\alpha+1+{\alpha\over 2})+\sqrt{D+1/4}+\sqrt{G+\alpha^2/4}
+\sqrt{F+\alpha^2/4}\bigg \}\omega\;.\eqno{(3.27)}$$
Since the degeneracy is usually associated with some symmetry in the
problem, it would be interesting to enquire about the extra symmetry
in the problem. For the  case $\alpha$ = 1 (apart from $\delta$ = C = 0),
the symmetry is clear. In such a case ,  the potential is
simply the spherically symmetric harmonic oscillator ,  with
the perturbations of the form $G /x^2$, $F /y^2$ and $C/z^2$. However,
in general, for any rational $\alpha$ ,  the symmetry is not obvious.

We thus have shown that by choosing
$V(r)$, $V(\theta)$ and $V(\phi)$ appropriately , the energy eigenvalues
and the eigenfunctions of the noncentral potential (1.1) can be written
down algebraically by using the results for the well known SIP. By
going through the list of the well known SIP ,  one can now immediately
write down E and $\Psi$ for a class of noncentral but separable  potentials. In
particular notice that another possible form for the shape invariant
$V(\phi)$ is [12] given by
$$V_2(\phi) = F {\rm  cosec}^2 \alpha\phi + G {\rm cot}
\alpha\phi\;.\eqno{(3.28)}$$
We may now use the relation [8] that if
$$W=-\,A\,{\rm cot}\,\alpha \phi\,-\,{B\over A}\;,\eqno(3.29)$$
then the corresponding (SIP) $V_{-}$ is given by
$$V_{-}=A(A-\alpha)\,{\rm cosec}^2\,\alpha\phi\,+\,2B\,{\rm cot}\,\alpha\phi
-A^2+{B^2\over A^2}\;.\eqno(3.30)$$
Hence the energy eigenvalues are
$$m^2_2 = (A+n\alpha)^2 - {B^2\over (A+n\,\alpha)^2}\;,\eqno(3.31)$$
where
$$A=(\alpha + \sqrt{\alpha^2+4 F })/2\;\;\;\;,\, B=G /2\;.\eqno(3.32)$$
Note that in unbroken supersymmetry, $A>0$ , and the negative root is
rejected.The corresponding eigenfunctions are found to be [8]
$$ K_2(\phi) = ({\rm sin}\alpha\phi)^{(n+s)} e^{a\alpha\phi}
P_n^{(-s-n-ia,-s-n+ia)}\,\,(i{\rm cot}\alpha\phi)\;,\eqno{(3.33)}$$
where
$$s = A/\alpha\; ,\, \lambda = {B\over \alpha^2}\;,\; a = {\lambda\over
\alpha(s+n)}\;.\eqno{(3.34)}$$

Similarly, the other possible forms for the shape invariant
$\theta$-dependent potential $V(\theta)$ are
$$V_2(\theta) = C\,{\rm cosec}^2\theta + D\,{\rm cot}\theta\;,\eqno{(3.35)}$$
and
$$V_3(\theta) = C\,{\rm cosec}^2\theta + D\,{\rm  cosec}\theta\,{\rm
cot}\theta\;.\eqno{(3.36)}$$
The  energy eigenvalues and eigenfunctions of the
Schr\"odinger eq.(3.11) for the potential $V_2(\theta)$ are [8]
$$l^2_2 = (A_1+n_1+{1\over 2})^2 - {D^2\over 4(A_1+n_1+{1\over 2})^2},\; A^2_1
=C + m^2\;,\eqno{(3.37)}$$
$$H_2(\theta) = ({\rm sin}\,\theta)^{(n_1+A_1+{1\over 2})} e^{a_1\theta}
P_{n_1}^{(-A_1-n_1-{1\over 2}-ia_1,-A_1-n_1-{1\over 2}+ia_1)} (i{\rm
cot}\,\theta)\;,\eqno{(3.38)}$$
where m is either $m_1$ or $m_2$ as given by eq.(3.13) or (3.31) respectively ,
and
$$a_1= {D\over 2(A_1+n_1+{1\over 2})}\;.\eqno{(3.39)}$$

In order to find the eigenvalues and the eigenfunctions for
the potential $V_3(\theta)$, note that if
$$W=-(A_1+{1\over 2})\,{\rm cot}\,\theta - B_1\,{\rm
cosec}\,\theta\;,\eqno(3.40)$$
then the corresponding (SIP) $V_{-}(\theta)$ is given by
$$V_{-}(\theta)=(A_1^2+B_1^2-{1\over 4})\;{\rm cosec}^2\theta+2A_1\,B_1
{\rm cot}\theta\;{\rm cosec}\theta\,-(A_1+{1\over 2})^2\;.\eqno(3.41)$$
Hence the the eigenvalues and the eigenfunctions of the
Schr\"odinger eq.(3.11) for the potential $V_3(\theta)$ are [4,6],
$$l^2_3 = (A_1+n_1+{1\over 2})^2\,\,,\eqno{(3.42)}$$
$$H_3(\theta) = ({\rm sin}\,{\theta\over 2})^{(A_1+B_1+{1\over 2})}
({\rm cos}{\theta\over 2})^{(A_1-B_1+{1\over 2})}P_{n_1}^{(A_1+B_1,A_1-B_1)}
 ({\rm cos}\theta)\;,\eqno{(3.43)}$$
where
$$m^2+C = A^2_1 + B^2_1, D = 2 A_1 B_1\;,\eqno{(3.44)}$$
with m being either $m_1$ or $m_2$ as given by Eqs.(3.13) or (3.31)
respectively.

Finally, the other possible form for the shape invariant r-dependent
potential V(r) is
$$V_2(r) = - {e^2\over r}+{\delta\over r^2}\;.\eqno{(3.45)}$$
Observe that [4] if
$$W={e^2\over{2(B_2+1)}}-{(B_2+1)\over r}\;,\eqno(3.46)$$
then the corresponding (SIP) $ V_{-}(r)$ is given by
$$V_{-}(r)=-{e^2\over r}+{B_2(B_2+1)\over
r^2}+{e^4\over{4(B_2+1)^2}}\;.\eqno(3.47)$$
Hence the energy eigenvalues  and the eigenfunctions for the potential
which follow from ref.[4,6] are
$$E^{(2)} = - {e^4\over 4(n_2+B_2+1)^2}\;,\eqno{(3.48)}$$
$$R_2(r) = y^{B_2+1} exp(-{1\over 2}y) L_{n_2}^{2B_2+1} (y)\;,\eqno{(3.49)}$$
where
$$ y = {e^2r\over (n_2+B_2+1)} ;\;\;(B_2 +{1\over 2})^2 = \delta +
l^2\;,\eqno{(3.50)}$$

with $l$ being $l_1$ or $l_2$ or $l_3$ depending on the form of $V(\theta)$.
We thus see that there are 2 $\phi$-dependent, 3 $\theta$-dependent and
2 r-dependent SIP. As a result, we can
immediately construct twelve different noncentral potentials by
taking various combinations of $V(r)$, $V(\theta)$ and $V(\phi)$ and in
each case we can write down the eigenvalues and the eigenfunctions
algebraically by using the well known results for the SIP as
mentioned above. The main results can be summarised as follows
\item {(i)} For six noncentral potentials, the energy eigenvalue are
given by the formula (3.25) and for the remaining six cases by the
formula (3.48) depending on if one is considering the oscillator
 or the Coulomb  potential. Ofcourse,the value of
$B_2$ will vary from potential to potential depending on the form of
$V(\theta)$ and $V(\phi)$ as explained above.
\item {(ii)} For all the 12 potentials, the eigenfunctions are given by
$$\psi(r,\theta,\phi) = R_i(r) H_j(\theta)K_q(\phi)$$
where i,q = 1,2 and j = 1,2,3, depending on the choice of the
potentials. For example, for the potential (1.1), $\psi =
R_1(r) H_1(\theta) K_1(\phi)$, where $R_1, H_1$ and $K_1$ are as
given by eqs.(3.26), (3.18) and (3.14) respectively.
\item {(iii)} In case $V(\theta)$ is
$V_1(\theta)$  or $V_3(\theta)$ ,  there is a degeneracy in the
spectrum for  $\delta = 0$ irrespective of the form of $V(r)$ and
$V(\phi)$. On the other hand, when $V(\phi)=V_1(\phi)$
there is a degeneracy in the spectrum if $C$ = 0 and $\alpha$ is a rational
number .However, if $V(\phi)
\,=V_2(\phi)$ and $V(\theta)\,=\, V_2(\theta)$  there
is no degeneracy in the spectrum
irrespective of the form of $V(r)$. Finally,
in case $V(\theta)$ = $V_1(\theta)$ or $V_3(\theta)$ , and
$V(\phi)$ = $V_1(\phi)$ ,  there is maximum degeneracy when  both C
and $\delta$ are zero and $\alpha$ is a rational number for either form
of $V(r)$. Degeneracy in the spectrum is usually associated with some
symmetry in the system , and it may be worthwhile to explore this
connection.
\item {(iv)} For the six noncentral potentials where V(r) is taken to be the
Coulomb potential, one also has a  continuous spectrum over and above
the discrete one . It would be interesting to obtain the phase
shifts for these noncentral potentials algebraically.
\vs

\nd{\bf IV. Conclusions}

In this paper, we have shown that there exist twelve noncentral but
separable potentials each of which have seven parameters and for
each of which the eigenvalues and the eigenfunctions may  be written down
in a closed form algebraically  using
the well known results for the shape invariant potentials. We find this
 remarkable . In this context, it is worth remembering that
even for central potentials, the most general solvable potentials are
those of Natanzon type with six parameters but where the potential is not
explicitly known [13].

Generalization of our technique to other noncentral potentials is
quite straight forward. For example, one could instead consider the
Schr\"odinger equation in the cylindrical coordinates and obtain other
examples of the noncentral potentials with seven parameters for which
the spectra can be written down  by using the well
known results for the SIP  $V(\phi)$, $V(\rho)$
and $V(z)$,  where $\rho=\sqrt{x^2+y^2}$. Other system of coordinates , or
even higher dimensions may similarly be considered.
\vs

\nd{\bf Acknowledgements}

This research was started  when R.K.B. was visiting the Institute of Physics .
The hospitality enjoyed  there is gratefully acknowledged. This
work was partially funded by N.S.E.R.C. of Canada.

\vfill
\eject
\noindent {\bf References}
\item {[1]} See for example ,  L. Schiff, Quantum Mechanics
(Mcgraw-Hill, New York, 1968); L. Landau and E. Lifshitz,
Quantum Mechanics (Pergamon, New York, 1977) ; R.Shankar ,
Principles of Quantum Mechanics (Pergamon , New York , 1977).
\item {[2]} E. Witten, Nucl. Phys. {\bf B185}, 513 (1981); F. Cooper
and B. Freedman, Ann. Phys. (N.Y) {\bf 146}, 262 (1983); For some
recent reviews on this field see A. Lahiri, P.K. Roy and B. Bagchi,
Int. J. Mod. Phys. {\bf A5}, 1383 (1990); A.Khare and U.P. Sukhatme, Phys. News
(India) {\bf 22}, 35 (1991).
\item {[3]} L. Gendenshtein, JETP Lett. {\bf 38}, 356 (1983).
\item {[4]} R. Dutt, A. Khare and U.P. Sukhatme, Ame. J. Phys. {\bf 56}, 163
(1988).
\item {[5]} R. Dutt, A. Khare and U.P. Sukhatme, Phys. Lett. {\bf
B181}, 295 (1986).
\item {[6]} J.W. Dabrowska, A. Khare and U.P. Sukhatme, Jour. of
Phys. {\bf A21}, L 195 (1988).
\item {[7]} F. Cooper, J. Ginocchio and A. Khare, Phys. Rev. {\bf
D36}, 2458 (1987).
\item {[8]} G. Levai, J. of Phys. {\bf A22}, 689  (1989).
\item {[9]} R. Dutt, A. Gangopadhyaya, A. Khare, A. Pagnamenta and
U.P. Sukhatme, Phys. Lett. {\bf A174}, 363 (1993).
\item {[10]} E. Schr\"odinger, Proc. Roy. Irish Acad. {\bf A46}, 9
(1940); L. Infeld and T.W. Hull, Rev. Mod. Phys. {\bf 23}, 21 (1951).
\item {[11]} See for example, P.M. Morse and H. Feshback, Methods of
Theoretical Physics, Vol.I, Chapter 5, Mcgraw Hill (N.Y) (1953).
\item {12]} The two other possible forms of SIP as
given by $V_3(\phi) = F \,{\rm cosec}^2\,\alpha\phi$ and $V_4(\phi) = F \,
 {\rm cosec}^2 \alpha\phi\, +\, G \,{\rm cot}\alpha\phi\;{\rm
cosec}\,\alpha\phi$  are not really independent as the
former is a special case $ (G  = 0)$ of Eq.(3.28) while the latter can be
reduced to  $V_1(\phi)$ as given by Eq.(3.9) , since $V_4(\phi)$ may
be written as $V_4(\phi)=({G+F\over 4}){\rm cosec}^2 ({\alpha\phi\over
2})+{(F-G)\over 4}{\rm sec}^2 ({\alpha\phi\over 2})$.
\item {[13]} P. Cordero and S. Salamo, J. Phys. {\bf A24}, 5200 (1991).
\vfill
\eject
\end

\end